\documentclass[12pt]{iopart}
\usepackage[T1]{fontenc}
\usepackage[utf8]{inputenc}
\usepackage[english]{babel}
\usepackage{graphicx,wrapfig}
\usepackage{epstopdf}
\usepackage{graphicx}
\usepackage{xcolor}
\usepackage{listings}

\begin{document}
	\title{Free energy expansion of the spin glass with finite connectivity for $\infty$ RSB}
	\author{Gioia Boschi\textsuperscript{1} and Giorgio Parisi \textsuperscript{2}}
	
	\address{\textsuperscript{1} Department of Mathematics, King's College London, Strand, London WC2R 2LS, United Kingdom\\
	\textsuperscript{2} Dipartimento di Fisica, Sapienza Università di Roma, Piazzale A. Moro 2, 00185 Rome, Italy\\
	}
	\ead{gioia.boschi@kcl.ac.uk, giorgio.parisi@roma1.infn.it}

\begin{abstract}
	In this paper, we investigate the finite connectivity spin-glass problem. Our work is focused on the expansion around the point of infinite connectivity of the free energy of a spin glass on a graph with Poissonian distributed connectivity: we are interested to study the first-order correction to the infinite connectivity result for large values or the connectivity $z$. The same calculations for one and two replica symmetry breakings were done in previous works; the result for the first-order correction was divergent in the limit of zero temperature and it was suggested that it was an artifact for having a finite number of replica symmetry breakings. In this paper we are able to calculate the expansion for an infinite number of replica symmetry breakings: in the zero-temperature limit, we obtain a well defined free energy. We have shown that cancellations of divergent terms occur in the case of an infinite number of replica symmetry breakings and that the pathological behavior of the expansion was due only to the finite number of replica symmetry breakings.
\end{abstract}

\vspace{2pc}
 
\section{Introduction}
In the last 40 years, a large amount of work has been dedicated to Spin Glass models with infinite connectivity, especially the very famous Sherrington-Kirkpatrick (SK) model \cite{sherrington1975solvable, mezard1987spin, morone2014replica}. Given also the technical difficulties, less attention has been dedicated to more realistic finite connection mean-field models ( \cite{mezard1985replicas, viana1985phase, orland1985mean, mezard1987mean}) with average connectivity $z$. The main problem common to finite connectivity systems is that the local fields are not Gaussian as infinite range models, but their distribution is a more complicated function. This implies that the order parameter is a function of the overlaps of any number of replicas instead of the overlap over only two of them as in the SK model. Consequently, in these models, it is difficult to find the exact free energy. 

A general solution for this problem has been proposed in \cite{mezard2001bethe}, \cite{mezard2003cavity} using the Bethe-Peierls cavity method. It is conjectured, but it remains unproven that this approach gives the correct result  (\cite{panchenko2016structure, parisi2017marginally,concetti2019properties} ). Perturbative expansions have been proposed near the infinite connectivity point (SK model) \cite{lai1990replica}, \cite{de1989replica}, \cite{goldschmidt1990replica}, \cite{parisi2002spin} (i.e. the $1/z$ expansion) and near the critical temperature \cite{mottishaw1987replica}.

 In this paper we consider spin glasses with Poisson distributed connectivity \cite{viana1985phase} and we study the large connectivity ($z$) expansion of the free energy for an infinite number of replica symmetry breakings (RSBs); at the end we specalize our formulae to the zero-temperature limit. In particular, we will calculate the first term of the expansion already investigated by Goldschmidt and De Dominicis \cite{de1989replica} and Parisi and Tria in \cite{parisi2002spin} respectively for 1RSB and 2RSB. De Dominicis and Goldschmidt found that the free energy expansion diverges at low temperatures for 1RSB and Parisi and Tria showed that for 2RSB the divergence is less pronounced. This indicates that the divergent behavior was due only to the finite number of RSBs. In this work, we show that when an infinite number of RSBs is considered, the divergent parts of the expansion cancel out leaving only a small residual divergence due to numerical errors.
 
This paper is organized as follows: in section (\ref{sec:model}) we describe the model and we write the expression for the first term of the free energy expansion as a function of $1/z$. This term will be composed of different integrals, that we compute in section (\ref{sec:divergence}). In the same section, we study the divergent behavior of these integrals for the temperature $T$ which goes to 0, showing analytically how we can reduce the divergences to a linear divergence in $\beta=1/T$. In the last section (\ref{sec:evaluation}) we finally combine the analytical results with the numerical evaluation to obtain an estimation of the first term of the free energy expansion, showing how its different divergent components cancel out. 

\section{The Large $z$ expansion for the Random connectivity model}\label{sec:model}
\subsection{The general formalism}
We consider a model with $N$ spins $\sigma_i=\pm 1$, with $i \in \{1...N\}$ interacting with random couplings and each one connected with $z_i$ other spins.  The coupling are defined on the edges of an Erd\"{o}s-Rény graph, with links drawn with probability $z/N$. The distribution of the connectivities ($z_i$) is Poissonian with mean $z$ in the limit for $N$ which goes to infinity. The couplings $J_{ij}$ distributed according to the following formula:
\begin{equation}
P(J_{ik})=\left( 1-\frac{z}{N}\right) \delta(J_{ik})+\frac{z}{N}\tilde{P}(J_{ik})\ .
\end{equation}
If two nodes of the graph are not connected the value of $J_{ij}$ is  null, while if they are connected the distribution of the values of $J_{ij}$ is  the following:
\begin{equation}
\tilde{P}(J_{ik})=\frac{1}{2}\left[ \delta\left(J_{ik}- \frac{1}{\sqrt{z}}\right)+\delta\left(J_{ik}+ \frac{1}{\sqrt{z}}\right)\right] \ \ \ \ \forall i,k \ .
\end{equation} 
This distribution has been chosen to be binary for simplicity, however we can obtain slightly more complicated expressions for Gaussian distributed $J_{ij}$.  Given the random structure of the lattice, the typical length of a loop is proportional to $\ln (N)$. Therefore in the infinite volume limit the graph approaches a Bethe lattice  that does not have finite size loops. For this reason the model belongs to the a mean field category, even if it cannot be  written in a simple way (as for the SK model) due to the difficulties related to the finite connectivity: for more explicit formulae see \cite{concetti2019properties}.

As usual we are interested in finding the average free-energy density defined as
\begin{equation}
f(z,\beta)=\lim_{N\to \infty}-\frac{1}{\beta N}\int dJ_{ik} P(J_{ik})\ln Z_J(z,\beta)\ ,
\end{equation}
with
\begin{equation}
 Z_J(z,\beta)=\sum_{\{\sigma\}}\exp\{-\beta H_J[\{\sigma\}]\}\ .
\end{equation}
It can be shown  that for $z\to \infty$ the mean free energy density becomes the SK one
\begin{equation}
\lim_{z\to\infty}f(z,\beta)=f_{SK}(\beta)\ .
\end{equation}
The computation of the free energy can be studied with the replica trick
\begin{equation}
f(n,z,\beta)=\lim_{N\to \infty}-\frac{1}{\beta N n}\ln \overline{Z^n}\ , \qquad
f(z,\beta)= \lim_{n\to 0}f(n,z,\beta)\ ,
\end{equation}
where $n$ is the number of replicas and $\overline{\ \cdot\ }$ denotes the average over the disorder. The partition function for the $n$ replicas is
\begin{eqnarray}
\nonumber \overline{Z^n}&=&\prod_{i<k}\int_{-\infty}^{+\infty}dJ_{ik}P(J_{ik})\sum_{\{\sigma\}}\exp(\beta J_{ik}\sum_{a}\sigma_i^a\sigma_k^a)\\ 
&=& \sum_{\{\sigma\}} \exp \{\frac{z}{N}\sum_{i<k}\left[ \cosh\left( \beta J_0\sum_a\sigma^a_i\sigma_k^a\right) -1\right] \}\ .
\end{eqnarray}
Following \cite{de1989replica} we expand the free energy in powers of $1/z$ around the infinity connectivity point in order to find the corrections to the SK free energy for large, but finite, connectivity.\\
 Expanding the hyperbolic cosine for a small argument we find a series in power of $\beta^2/z$. If we neglect terms of order $(\beta^2/z)^3$ we find:
\begin{eqnarray}
\overline{Z^n}&=&  \sum_{\{\sigma\}} \exp \{\frac{z}{N}\sum_{i<k}[\frac{\beta^2}{2z}(\sum_a\sigma_i^a\sigma_k^a)^2+\frac{\beta^4}{24z^2}(\sum_a\sigma_i^a\sigma_k^a)^4]     \}=\\
&=& \nonumber \sum_{\{\sigma\}}\exp \{\frac{1}{N}\sum_{i<k}[\frac{\beta^2}{2}(\sum_{a,b}\sigma_i^a\sigma_k^a\sigma_i^b\sigma_k^b)+\frac{\beta^4}{24z}(\sum_{a,b,c,d}\sigma_i^a\sigma_k^a\sigma_i^b\sigma_k^b\sigma_i^c\sigma_k^c\sigma_i^d\sigma_k^d)]     \}\ .
\end{eqnarray}

For $z$ which goes to infinity only the first term of the sum survives, which corresponds to the SK free energy $f_{SK}$. As a consequence, we can write the free energy expansion as
\begin{equation}\label{2}
f(n, z,\beta)=f_{SK}(n, \beta)+\frac{1}{z} f_1 + o\left(\frac{1}{z^2}\right)\ ,
\end{equation}
where 
\begin{equation}\label{6}
\beta f_1= \frac{\beta^4}{24}+\frac{\beta^4}{3n}\sum_{a<b}q_{ab}^{2}-\frac{\beta^4}{2n}\sum_{a<b<c<d}q_{abcd}^{2}\ ,
\end{equation}
with
\begin{equation}
q_{abcd}=\frac{1}{N}\sum_{i}\sigma_i^a\sigma_i^b\sigma_i^c\sigma_i^d\ .
\end{equation}
The quantity above is the overlap among 4 replicas, $q_{abcd}$, which is the generalization of the one for two replicas $q_{ab}$.

As mentioned before, the expression for the second term of the free energy expansion $f_1$ (Eq. \ref{6}) has been firstly derived by De Dominicis and Goldschmidt in \cite{de1989replica}, which also evaluated it in 1RSB, while Tria and Parisi \cite{parisi2002spin} evaluated it in 2RSB. In this work we will show how to calculate it explicitly for an infinite number of RSBs. In order to do so, we need to work out the sums over the replica indices which appear in it.
\subsection{Explicit computation of first order }
Following the calculation of \cite{parisi2002spin} we can obtain Eq. \ref{6}. First of all, we write the sums in the following way:
\begin{eqnarray}
\sum_{a,b}\sigma^a\sigma^b&=&\sum_{a\ne b}\sigma^a\sigma^b+n\\
\sum_{a,b,c,d}\sigma^a\sigma^b\sigma^c\sigma^d&=&\sum_{a\ne b\ne c\ne d}\sigma^a\sigma^b\sigma^c\sigma^d -8\sum_{a\ne b}\sigma^a\sigma^b-2n\ ,
\end{eqnarray}
and we use the standard Gaussian formula  :
\begin{equation} 
\exp\left(\frac{1}{2}\lambda^2 a\right)=\frac{1}{\sqrt{2\pi a}}\int dx \exp \left(-\frac{x^2}{4a}+\lambda a x \right)\ . 
\end{equation}
We set $\lambda=\sum_i\sigma_i^a\sigma_i^b/N$ and $ a=N\beta^2$ for the $\beta^2$ term and $\lambda=\sum_i\sigma_i^a\sigma_i^b\sigma_i^c\sigma_i^d/N$ and $a=\beta^4/12z $ for the $\beta^4$ one. Recalling that
\begin{equation}
\sum_{a<b<c<d}=\frac{1}{4!}\sum_{a\ne b\ne c\ne d}\ ,\qquad 
\sum_{a<b}=\frac{1}{2!}\sum_{a\ne b}\ ,
\end{equation}
we arrive to the following integral representation:
\begin{equation}
\overline{Z^n}=\int \prod_{a<b}dQ_{ab}\prod_{a<b<c<d}dQ_{abcd}\exp(-Nf[Q])\ ,
\end{equation}
with:
\begin{eqnarray}
f[Q]&=&-\frac{\beta^2}{4}+\frac{\beta^4}{24z}+\frac{\beta^2}{2}\sum_{a<b}Q_{ab}^2-\frac{\beta^4}{3z}\sum_{a<b}Q_{a,b}^2+\\ \nonumber
&+&\frac{\beta^4}{2z}\sum_{a<b<c<d}Q_{abcd}^2-\ln\{\mbox{Tr}_\sigma\exp(G[Q,\sigma])\},\\
G[Q,\sigma]&=&\beta^2\sum_{a<b}Q_{ab}\sigma_a\sigma_b-\frac{2\beta^4}{3z}\sum_{a<b}Q_{ab}\sigma_a\sigma_b+\frac{\beta^4}{z}\sum_{a<b<c<d}\sigma_a\sigma_b\sigma_c\sigma_d\ ,
\end{eqnarray}
where we use the notation
\begin{equation}
\mbox{Tr}_\sigma\exp(G[Q,\sigma])=\left(\prod_{a=1,n}\sum_{\sigma_a } \right)\exp(G[Q,\sigma])\ .
\end{equation}

At the and of the day we find:
\begin{equation}\label{5}
\beta f_1=  \frac{\beta^4}{24}+\frac{\beta^4}{3n}\sum_{a<b}q_{ab}^{(0)^2}-\frac{\beta^4}{2n}\sum_{a<b<c<d}q_{abcd}^{(0)^2}\ .
\end{equation}

Generally speaking the finite connectivity model is particularly hard to study because of the appearance of overlaps among an increasing number of replicas as the order of the free energy expansion grows. Concerning the first order correction, we will have only the following two overlaps:
\begin{eqnarray}
q_{ab}&=& q_{ab}^{(0)}+\frac{1}{z}q_{ab}^{(1)}+...\ ,\\
q_{abcd}&= & q_{abcd}^{(0)}+\frac{1}{z}q_{abcd}^{(1)}+...\ ,
\end{eqnarray} 
If we are interested to compute only the leading corrections, we can use their asymptotic expression: 
\begin{eqnarray}
q_{ab}^{(0)}&=&\frac{\mbox{Tr}_\sigma\exp[\beta^2\sum_{r<s}q_{rs}^{(0)}\sigma_r\sigma_s]\sigma_a\sigma_b} {\mbox{Tr}_\sigma\exp[\beta^2\sum_{r<s}q_{rs}^{(0)}\sigma_r\sigma_s]}\equiv\langle\sigma_a\sigma_b\rangle_Q \ ,\\
q_{abcd}^{(0)} &=&\frac{\mbox{Tr}_\sigma\exp[\beta^2\sum_{r<s}q_{rs}^{(0)}\sigma_r\sigma_s]\sigma_a\sigma_b\sigma_c\sigma_d} {\mbox{Tr}_\sigma\exp[\beta^2\sum_{r<s}q_{rs}^{(0)}\sigma_r\sigma_s]}\equiv \langle \sigma_a\sigma_b\sigma_c\sigma_d\rangle_Q\ ,
\end{eqnarray}
and $\langle \ \cdot\ \rangle _Q$ is the average on the single site SK Hamiltonian using $Q=q^{(0)}$.
\subsection{Continuous replica symmetry breaking} 
\begin{figure}[h]
\centering
\includegraphics[scale=0.3]{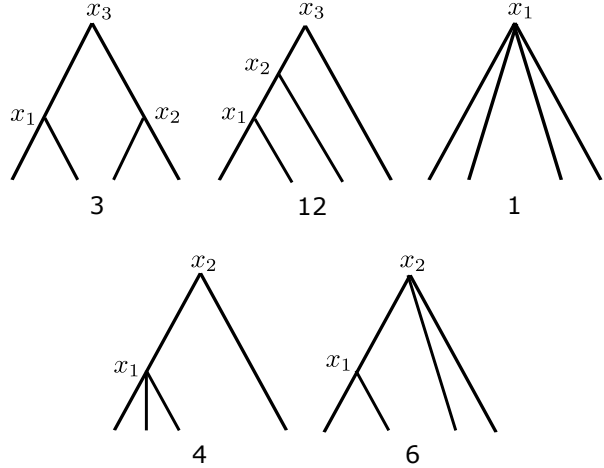}
\caption{Ultrametric trees for the 4 replicas overlap, by G. Parisi e F. Tria \cite{parisi2002spin}.}\label{3}
\end{figure}

For an infinite number of symmetry breakings we can express the sums in Eq. \ref{5} as integrals. We have the following expression for the two indices overlap:
\begin{equation}
\lim_{n\to 0}\frac{1}{n}\sum_{ab}q_{ab}^2=-\int_{0}^{1}q^2(x)dx\ ,
\end{equation}
with
\begin{equation}\label{q(x)}
q(x)=q_{ab|_{a \wedge b= x}}\ ,
\end{equation}
where $ a \wedge b= x$  means that $a$ and $b$ belong to symmetry breakings blocks of distance $x$.
Concerning the sum on four indices, we will need to consider the four indices overlap, which is a function of six variables
\begin{equation}
q(x_{12},x_{23},x_{34},x_{13},x_{14},x_{24})\ .
\end{equation}
It can be shown via ultrametric properties \cite{mezard1985microstructure} that only 3 out of the 6 variables are independent. Mezard and Yedidia \cite{MezardYedidia} showed that the sum on the four indices can be found via a diagrammatic representation with five trees (figure \ref{3}), each one with at maximum of three nodes. For each tree we need to keep in consideration also a multiplicative factor $x^{s-2}(s-2)!$ for each vertex, where $s$ is the number of lines entering in the vertex. The resulting sum will be
\begin{eqnarray}\label{17}
-\frac{24}{n}\sum_{a<b<c<d}q_{abcd}^2=3\int_{0}^{1}dx_3\int_{x_3}^{1}dx_2\int_{x_3}^{1}dx_1q_1(x_1,x_2,x_3)^2+\\\nonumber
+12\int_{0}^{1}dx_3\int_{x_3}^{1}dx_2\int_{x_2}^{1}dx_1q_2(x_1x_2x_3)^2
+\int_{0}^{1}dx_12x_1^2q_3^2(x_1)+\\\nonumber
+4\int_{0}^{1}dx_2\int_{x_2}^{1}dx_1x_1q_4(x_1,x_2)^2+6\int_{0}^{1}dx_2\int_{x_2}^{1}dx_1q_5(x_1,x_2)^2\ ,
\end{eqnarray}
where  the functions $q_i$  are defined in \ref{appendix_qi}.

The next step is the computation of the functions $q_i$ to solve Eq. \ref{17}. For user convenience let us recall the known results on the solution of the SK model.

Using the variational method of Sommers and Dupont \cite{sommers1984distribution} and the results of \cite{mezard1985microstructure} we can introduce here the functions $P(x,z|x_0,z_0)$ and $m(x,z)$. Differential equations which characterize these quantities can be found imposing the stationarity to the Parisi's free energy \cite{parisi1980sequence} with respect to $P(x,z)$, $P(1,z)$, $\phi(x,z)$, $\phi(0,y)$ and $q(x)$, where $\phi(x,z)$ is the solution of
\begin{equation}
\dot{\phi}(x,z)=-\frac{\dot{q}(x)}{2}[\phi''(x,z)+\beta x\phi'(x,z)^2]\ ,
\end{equation}
with boundary conditions
\begin{equation}
\phi(1,z) =\beta^{-1}\log(2\cosh \beta z) .
\end{equation}
In our notation the derivatives with respect to $x$ are denoted by a dot, while those respect to $z$ are denoted by a prime. With the variational approach we thus find
\begin{eqnarray}
q(x) &=&\int dz P(x,z)m^2(x, z) \label{q(x)_2}\ ,\\
\dot{m}(x, z)&=&-\frac{\dot{q}(x)}{2}[m''(x,z)+2\beta x m(x,z)m'(x,z)]\ ,\\
\dot{P}(x,z) &=&\frac{\dot{q}(x)}{2} [P''(x,z)-2\beta x [m(x,z)P(x,z)]']\ ,
\end{eqnarray}
with initial conditions
\begin{eqnarray}
m(1,z) &=& \tanh(\beta z)\ ,\\
P(0,z) &=& \delta(z)\ .
\end{eqnarray}
 We can write the  equations above as stochastic differential equations \cite{mezard1985microstructure, parisi1984turbulence}. To do so we introduce the auxiliary function $z_\eta(x)$, for $0\le x \le 1$. It satisfies the stochastic differential equation
\begin{equation}\label{z_eta}
\dot{z_\eta}(x)=\eta(x)\sqrt{\dot{q}(x)} - \beta x m(x,z)\dot{q}(x)\qquad  {z_\eta}(x),
\end{equation}
where $\eta(x)$ is a white noise with
\begin{equation}
\overline{\eta(x)\eta(x')}=\delta(x,x')\ ,
\end{equation}
and $\overline{\ \ \cdot\ \ }$ is the average over the noise $\eta$. The quantity $q(x)$ is the order parameter defined in Eq. \ref{q(x)} and the function $m(x,z)$ satisfies:
\begin{equation}\label{m(x,z)}
m(x,z)=\overline{\tanh(\beta z_\eta(x=1))}\ ,
\end{equation}
where the average is done over all the trajectories $z_\eta$ which go from $z(x)$ to $z(x=1)$ and the boundary conditions of Eq. \ref{z_eta} are $z_\eta(x)=z(x)$. This variational form has been recently used by \cite{auffinger2015parisi}
As a consequence, the two indices overlap in Eq. \ref{q(x)_2} can 
be calculated as 
\begin{equation}
	q(x)=\overline{m^2(x,z_\eta(x))} =\int dz P(x,z;0,0) m^2(x,z) \, .
\end{equation}

\begin{figure}[h]
	\centering
	\includegraphics[scale=0.3]{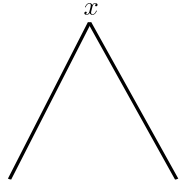}
	\caption{Utrametric tree for the two indices overlap, by G. Parisi e F. Tria \cite{parisi2002spin}}\label{1} 
\end{figure}

In other words quantity $q(x)$ can 
be represented as the average of $m^2(x,z)$ using as weight the probability  $P(x,z;0,0)$ for having a trajectory $z_\eta(x)$ taking the value $z$ at $x$, conditioned to have the value 0 at 0 (in the following we denote by P$(x,z;x_s,z_s)$ the conditional probability for having a trajectory that takes the value $z_s$ at $x_s$ and take the value $z$ at $x$ with $x>x_s$).  We recall that $m(x,z)$ is the average value of the magnetization of the trajectories  that taking the value $z$ at $x$ end up in $x=1$.  We find convenient represent the quantity $q(x)$ by the tree diagram shown in Figure \ref{1}.

In the same way we can rewrite all the integrals in Eq. \ref{17}, considering the five possible ways in which the replicas can be organized (represented by the tree diagrams in Figure \ref{3}):
\begin{eqnarray}\nonumber \label{q1}
q_1(x_1,x_2,x_3)=\int dz_1dz_2dz_3 P(x_3,z_3|0,0)P(x_2,z_2|x_3,z_3)\\P(x_1,z_1|x_3,z_3)
m^2(x_2,z_2)m^2(x_1,z_1)\ ,
\end{eqnarray}
\begin{eqnarray}\label{q2}
q_2(x_1,x_2,x_3)=\int dz_1dz_2dz_3 P(x_3,z_3|0,0)P(x_2,z_2|z_3,x_3)\\
P(x_1,z_1|z_2,x_2)m(x_3,z_3)m(x_2,z_2)m^2(x_1,z_1)\ ,
\end{eqnarray}
\begin{equation}
q_3(x_1)=\int dz_1 P(x_1,z_1|0,0)m^4(x_1,z_1)\ ,
\end{equation}
\begin{equation}
q_4(x_1,x_2)=\int dz_1dz_2P(x_2,z_2|0,0)
P(x_1,z_1|x_2,z_2)m(x_2,z_2)m^3(x_1,z_1)\ ,
\end{equation}
\begin{equation}
q_5(x_1,x_2)=\int dz_1dz_2P(x_2,z_2|0,0)P(x_1,z_1|x_2,z_2)m^2(x_2,z_2)m^2(x_1,z_1)\ .
\end{equation}

If we now introduce the sum over   the four indices we find:
\begin{eqnarray}\label{13}
-\frac{24}{n}\beta^3\sum_{a<b<c<d}q_{abcd}^2=\left[ 3\int_{0}^{\beta}dy_3\int_{y_3}^{\beta}dy_2\int_{y_3}^{\beta}dy_1q_1(y_1,y_2,y_3)^2+\right. \\\nonumber
+12\int_{0}^{\beta}dy_3\int_{y_3}^{\beta}dy_2\int_{y_2}^{\beta}dy_1q_2(y_1,y_2,y_3)^2
+\int_{0}^{\beta}dy_12y_1^2q_3^2(y_1)+\\\nonumber
\left. +4\int_{0}^{\beta}dy_2\int_{y_2}^{\beta}dy_1 y_1q_4(y_1,y_2)^2+6\int_{0}^{\beta}dy_2\int_{y_2}^{\beta}dy_1q_5(y_1,y_2)^2\right]\ .
\end{eqnarray}

We have written all the integrals as functions of $y=\beta x$. We choose this notation in order to keep track of all the $\beta$ factors that appear in the integrals, such that we can check if there appear divergences in the $\beta \to \infty$ limit.
Now that we have all the explicit expressions needed for $f_1$, we need to calculate them and take the limit for $\beta$ which goes to $\infty$.  
\section{Behaviour  of the first-order correction at large $\beta$}\label{sec:divergence}

Using the formalism of the previous section we can write the formal expansion
\begin{equation}
f(z,\beta)=\sum_k f_k(\beta) z^{-k} \,.
\end{equation}
What happens in the limit $\beta \to \infty$? In the best of the possible words we have
\begin{equation}
f(z,\infty)=\sum_k f_k(\infty) z^{-k} \qquad f_k(\infty) =\lim_{\beta \to \infty} f_k(\beta) \,. 
\end{equation}

However, the previous relation is not granted and the existence of the limits is unclear. It has been remarked in \cite{boettcher2003numerical} that in the case of the random regular graph with binary $J$ the values of the energy have oscillations for even-odd values of $z$ that apparently do not vanish exponentially so that the $1/z$ expansion should contain also oscillating terms. In the Poisson case considered here the values of $z$  may be non integer and there is no sign of
oscillations \cite{boettcher2003numerical}.

Here our aim is more limited: we investigate the existence of the limit
\begin{equation}
\lim_{\beta \to \infty} f_k(\beta) 
\end{equation}
in the case $k=1$. It is quite possible that for Poisson variables the limit $z\to \infty$ and $\beta \to \infty$ can be exchanged, however this is a delicate issue that will be not addressed here.

The behavior of $f_1(\beta)$ at large $\beta$ is interesting given that $f_1(\beta)$ diverges in the replica symmetric case and when the replica symmetry is broken at a few steps. It has been conjectured that this quantity has a finite limit in the case of continuous replica symmetry breaking. Our numerical estimates do confirm these expectations.
Let us consider $f_1(\beta)$; it contains three terms:
\begin{equation}
 f_1(\beta)= \frac{\beta^3}{24}+\frac{\beta^3}{3n}\sum_{a<b}q_{ab}^{(0)^2}-\frac{\beta^3}{2n}\sum_{a<b<c<d}q_{abcd}^{(0)^2}\ ,
\end{equation}
we want to check that there are no divergences in $f_1$ when $\beta$ goes to $\infty$. It is well known from \cite{de1989replica} and \cite{parisi2002spin} that for $\beta \to \infty$ the free energy expansion is divergent for 1RSB and remain divergent for 2RSB. The only way to have a non divergent expansion is to calculate it for $\infty$-RSB. 
 As we can see from Eq. \ref{13}, the integrals that we need to evaluate have an upper limit equal to $\beta$, which goes to infinity for $T\to 0$. Therefore, we will see two kind of divergences: the first one due to the multiplicative $\beta$ factors, the second due to the infinite interval of integration. In this section we want to see if the terms in $\beta$ and $\beta^3$, which diverges for $\beta \to \infty$, cancel out. 
\subsection{$\beta^3$ divergences}
The behaviour of $q(y)$ for $\beta \to \infty$ has been already studied by Pankov in \cite{pankov2006low}. The author showed that $q(x)$ can be written as
\begin{equation}
q(x)=1-c(\beta x)^{-2}-c_1(\beta x)^{-2}x^{\lambda}\ ,
\end{equation}
and becomes exact in the limit of the scaling regime ($\beta \to \infty$ and $x<<1$), where the factors $c,c_1, \lambda$ were computed by the author. Hence in our case we will have:
\begin{equation}
q(y) = 1 - \frac{c}{y^2} + o(y^{-2})\ ,
\end{equation}
in function of $y$. For the multiple variables overlap the situation is more complicated. We know that if all the variables go to infinity, the overlap approaches 1. In the first stage of the calculation we will assume that the corrections to this limit approach zero sufficiently rapidly. If this happen, in order to assure the \emph{superficial} convergence of the integrals, we just need to subtract 1 to the argument. We can write
\begin{equation}\label{integrals-1}
 -\frac{24}{n}\sum_{a<b<c<d}q_{abcd}^2=I_1+A_{1}\ ,
\end{equation}
where we divided the terms in two parts, one superficially convergent ($I_1$) and one divergent ($A_{1}$). The superficially convergent part is:
\begin{eqnarray}\label{convergent}
I_{1}&=&\frac{1}{\beta^3}\left[ 3\int_{0}^{\beta}dy_3\int_{y_3}^{\beta}dy_2\int_{y_3}^{\beta}dy_1[q_1(y_1,y_2,y_3)^2-1]+\right. \\\nonumber
&\ &+12\int_{0}^{\beta}dy_3\int_{y_3}^{\beta}dy_2\int_{y_2}^{\beta}dy_1[q_2(y_1,y_2,y_3)^2-1]+\\\nonumber
&\ &+\int_{0}^{\beta}dy_12y_1^2[q_3^2(y_1)-1]+
4\int_{0}^{\beta}dy_2\int_{y_2}^{\beta}dy_1y_1[q_4(y_1,y_2)^2-1]+\\
&\ &\left. +6\int_{0}^{\beta}dy_2\int_{y_2}^{\beta}dy_1[q_5(y_1,y_2)^2-1]\right]\ ,
\end{eqnarray}
while the non  convergent part is:
\begin{eqnarray}
A_{1}=\frac{1}{\beta^3}\left[ 3\int_{0}^{\beta}dy_3\int_{y_3}^{\beta}dy_2\int_{y_3}^{\beta}dy_1+\right. 
12\int_{0}^{\beta}dy_3\int_{y_3}^{\beta}dy_2\int_{y_2}^{\beta}dy_1+\\ \nonumber
+\int_{0}^{\beta}dy_12y_1^2+
\left. 4\int_{0}^{\beta}dy_2\int_{y_2}^{\beta}dy_1y_1+6\int_{0}^{\beta}dy_2\int_{y_2}^{\beta}dy_1\right]\ .
\end{eqnarray}
If we multiply $A_1$ for the proper factor from Eq. \ref{13} we find that the contribution of the non convergent part to the finale results is  :
\begin{equation} 
\frac{\beta^3}{48}A_{1}=\frac{1}{8}\beta^3\ ,
\end{equation}
that  is clearly divergent for $\beta \to \infty$. 

We can do the same for the two indices sum:
\begin{equation}
-\lim_{n\to 0}\frac{1}{n}\sum_{ab}q_{ab}^2\equiv\frac{1}{\beta}\int_{0}^{\beta}q^2(y)dy=\frac{1}{\beta}\int_{0}^{\beta}[q^2(y)-1]dy + A_{2}=I_2+A_{2}\ ,
\end{equation}
with
\begin{equation}\label{convergent_2}
I_2=\frac{1}{\beta}\int_{0}^{\beta}[q^2(y)-1]dy
\end{equation}
convergent and the term $A_{2}=1$ which contributes to the $\beta^3$  divergent term. The divergent part, multiplied by the factor from Eq. \ref{13}, gives:
\begin{equation}
\frac{\beta^3}{6}A_{2}=\frac{\beta^3}{6}\ .
\end{equation}

The last term to analyze is $\beta^3/24$, which is divergent at zero temperature. If we sum together all the divergent terms, we can see that they cancel out, solving the problem of the divergences in $\beta^3$:
\begin{equation}
\frac{\beta^3}{24}-\frac{\beta^3}{6}+\frac{\beta^3}{8}=0\ .
\end{equation}

\subsection{$\beta^2$ divergences}
The divergences check is not finished, we have in fact to check the  \emph{superficially} convergent terms.  We shall see that these terms are only convergent if we do a superficial analysis, but they are divergent as $\beta$. 

Let us consider to the following contribution to $f_1(\beta)$:
\begin{equation}
 G_1(\beta)\equiv -\frac{\beta^3}{3}I_2(\beta)+\frac{\beta^3}{2}I_1(\beta)\ .
\end{equation} 
Using the asymptotic behaviour of $q(y)$ (i.e. $1-c/y^2$) we find that $I_2(\beta)$ contains both a term proportional to $\beta^{-1}$ and a term equal to $-c\beta^{-2}$. These two terms multiplied py $\beta^3$ give a term proportional to $\beta^2$ and $\beta$. Similar terms are present also in $I_1(\beta)$. We shall see now the cancellation of the $\beta^2$ out.
\subsubsection{Calculations for the first integral}
Let us analyse the behaviour of the first integral appearing in $I_1$ (Eq. \ref{convergent}):
\begin{eqnarray}
\int_{0}^{\beta}dy_3\int_{y_3}^{\beta}dy_2\int_{y_3}^{\beta}dy_1(q_1(y_1,y_2,y_3)^2-1)\ .
\end{eqnarray}
Before analysing the function $q_1(y_1,y_2,y_3)$, we can manipulate it in order to simplify the rest of the calculations. As we can see from the ultrametric tree in Fig. \ref{3}, the argument of the integral is symmetric under the exchange of $y_1$ and $y_2$, so we can write:
\begin{eqnarray}
\int_{0}^{\beta}&dy_3\int_{y_3}^{\beta}dy_2\int_{y_3}^{\beta}dy_1(q_1(y_1,y_2,y_3)^2-1) = \\ \nonumber
&=2\int_{0}^{\beta}dy_3\int_{y_3}^{\beta}dy_2\int_{y_2}^{\beta}dy_1(q_1(y_1,y_2,y_3)^2-1)\ .
\end{eqnarray}
As in Eq. \ref{q1}, the function $q_1(y_1,y_2,y_3)$ can be written as
\begin{eqnarray}
q_1(y_1,y_2,y_3)=\int dz_1dz_2dz_3 P(y_3,z_3|0,0)P(y_2,z_2|y_3,z_3)\\P(y_1,z_1|y_2,z_2)\nonumber
m^2(y_2,z_2)m^2(y_1,z_1)\ ,
\end{eqnarray}
with the conditions $y_2>y_3$ and $y_1>y_2$. For $y_1$ going to infinity:
\begin{equation}\nonumber
\centering y_1 \to \infty \ \ \ \ \ \ \ \ \ \ m^2(y_1,z_1)\to 1\ ,
\end{equation}
the expression for $q_1(y_1,y_2,y_3)$ becomes:
\begin{eqnarray}\label{q_1}
q_1(y_1,y_2,y_3)=\int dz_2dz_3 P(y_3,z_3|0,0)P(y_2,z_2|y_3,z_3)m^2(y_2,z_2)\ ,
\end{eqnarray}
where:
\begin{eqnarray}
\int dz_3 P(y_3,z_3|0,0)P(y_2,z_2|y_3,z_3)&=& P(y_2,z_2|0,0)\ ,\nonumber \\
\int dz_2P(y_2,z_2|0,0)m^2(y_2,z_2)&=& q(y_2)\ .
\end{eqnarray}
Therefore the divergent part of the first diagram is:
\begin{equation}
\int_{0}^{\beta} dy_3\int_{y_3}^{\beta}dy_2\int_{y_2}^{\beta}dy_1 [q(y_2)-1]\ .
\end{equation}
With a similar reasoning we can find the divergent part of all the other diagrams. After having multiplied them for the appropriate factors and summed them together we obtain:
\begin{eqnarray}
A_{3}&\equiv&\frac{1}{\beta^3}\left[ 6\int_{0}^{\beta} dy_3\int_{y_3}^{\beta}dy_2\int_{y_2}^{\beta}dy_1 [q(y_2)-1]+\right.\\
 \nonumber &\ &+  12\int_{0}^{\beta} dy_3\int_{y_3}^{\beta}dy_2\int_{y_2}^{\beta}dy_1 [q^2(y_3)-1]+ \\
 \nonumber 
&\ &+ 4\left. \int_{0}^{\beta} dy_2\int_{y_2}^{\beta}dy_1y_1 [q^2(y_2)-1]+6\int_{0}^{\beta} dy_2y_2\int_{y_2}^{\beta}dy_1 [q^2(y_2)-1]\right] \\\nonumber
&=& -8\frac{1}{\beta^3}\int_{0}^{\beta}dyy^2[q^2(y)-1]+8\frac{1}{\beta}\int_{0}^{\beta}dy[q^2(y)-1]\\
\nonumber &=& -8\frac{1}{\beta^3}\int_{0}^{\beta}dyy^2[q^2(y)-1]+8I_2\ ,
\end{eqnarray}
so the remaining part of $G_1(\beta)$ is:
\begin{eqnarray}\label{19}
 f_1&=&-\frac{\beta^3}{6}I_2+\frac{\beta^3}{48}[I_1-A_{3}]+\frac{\beta^3}{48}A_{3}=\\ \nonumber
 &=&-\frac{\beta^3}{6}I_2+\frac{\beta^3}{48}[I_1-A_{3}]- \frac{\beta^3}{6}\int_{0}^{\beta}dyy^2[q^2(y)-1]+\frac{\beta^3}{6}I_2=\\\nonumber
&=&\frac{\beta^3}{48}[I_1-A_{3}]-\frac{1}{6} \int_{0}^{\beta}dyy^2[q^2(y)-1]\ .
\end{eqnarray}
One could hope that everying is now convergent. Unfortunately at this point there is another kind of divergence not yet analysed. If we take the last term of the sum in Eq. \ref{19}, we can show that this diverges linearly as $\beta$ goes to infinity, in fact:
\begin{equation}
q(y)\sim 1-\frac{c}{y^2}\ ,
\end{equation}
so:
\begin{equation}
\int_{0}^{\beta}dyy^2[q^2(y)-1]\sim\int_{0}^{\beta}dyy^2\frac{c}{y^2}=\int_{0}^{\beta}c dy=\beta c\ .
\end{equation}
We can argue that the multivariable overlap has the same trend in $y$ and that the other integrals have the same divergence problem. Therefore we can expect a reciprocal cancellation of the linear divergences in $\beta$ which come out from these integrals. However this last claim can not be verified analytically, so we used numerical techniques to prove it.
\section{Evaluation of $f_1$ at $T=0$ and conclusions}
\label{sec:evaluation}

Given that is not possible to further simplify the problem to check analytically if a reciprocal cancellation of the terms leads to a convergent $f_1$, we will evaluate numerically (details can be found in \ref{app:num_ev}) the different integrals appearing in Eq. (\ref{19}) which can be written in the following way:
\begin{eqnarray}\label{20}
f_1&=&\frac{1}{48}\left[ 6\int_{0}^{\beta}dy_3\int_{y_3}^{\beta}dy_2\int_{y_2}^{\beta}dy_1[q_1^2(y_1,y_2,y_3)-q^2(y_2)]+\right. \\\nonumber
&\ &+ 12\int_{0}^{\beta}dy_3\int_{y_3}^{\beta}dy_2\int_{y_2}^{\beta}dy_1[q_2^2(y_1,y_2,y_3)-q^2(y_3)]+\\\nonumber
&\ &+\int_{0}^{\beta}dy_12y_1^2[q_3^2(y_1)-1]+
4\int_{0}^{\beta}dy_2\int_{y_2}^{\beta}dy_1y_1[q_4^2(y_1,y_2)-q^2(y_2)]+\\
\nonumber &\ &+ \left. 6\int_{0}^{\beta}dy_2y_2\int_{y_2}^{\beta}dy_1[q_5^2(y_1,y_2)-q^2(y_2)]\right] -\frac{1}{6}\int_{0}^{\beta}dyy^2[q^2(y)-1]\\
&=& C1 + C2 + C3 + C4 + C5 + C6\nonumber\ .
\end{eqnarray}
 In order to evaluate the magnetizations appearing in the integrals (e.g. the ones in Eq. \ref{q_1}) we generated trajectories of the following form
\begin{equation}\label{15}
z_{i+1}=z_i+\sqrt{\epsilon}\eta(q_i)-y(q_i)m(q_i,z_i)\epsilon\ ,
\end{equation}
with $\epsilon=q_{i+1}-q_{i}$ and $q_i$ which goes from 0 to $q_{MAX}$, following the branching point of the trees in Fig. \ref{3} (see \ref{app:num_ev}).
\begin{figure}
	\centering
	\includegraphics[scale=0.4]{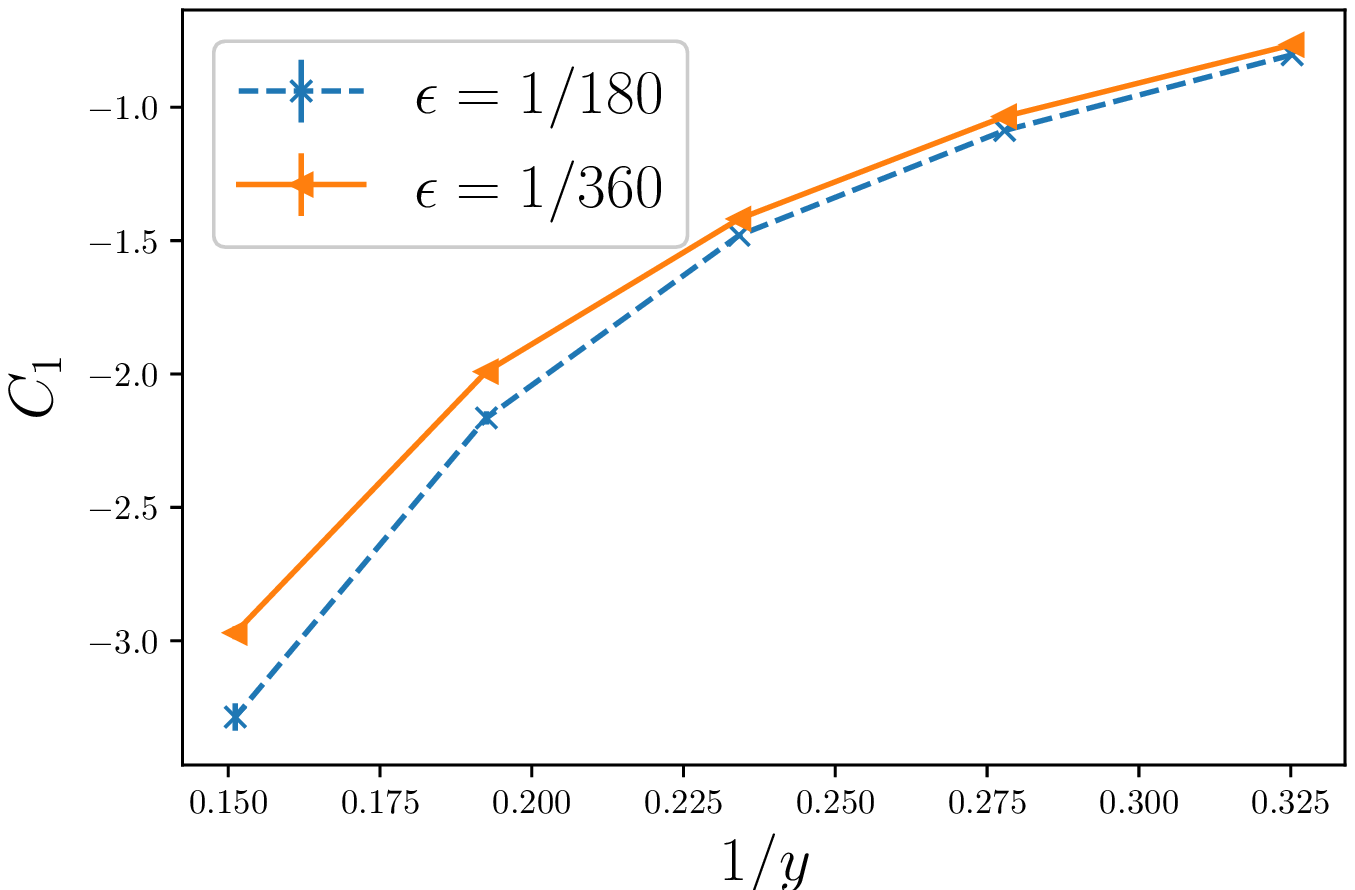}
	\includegraphics[scale=0.4]{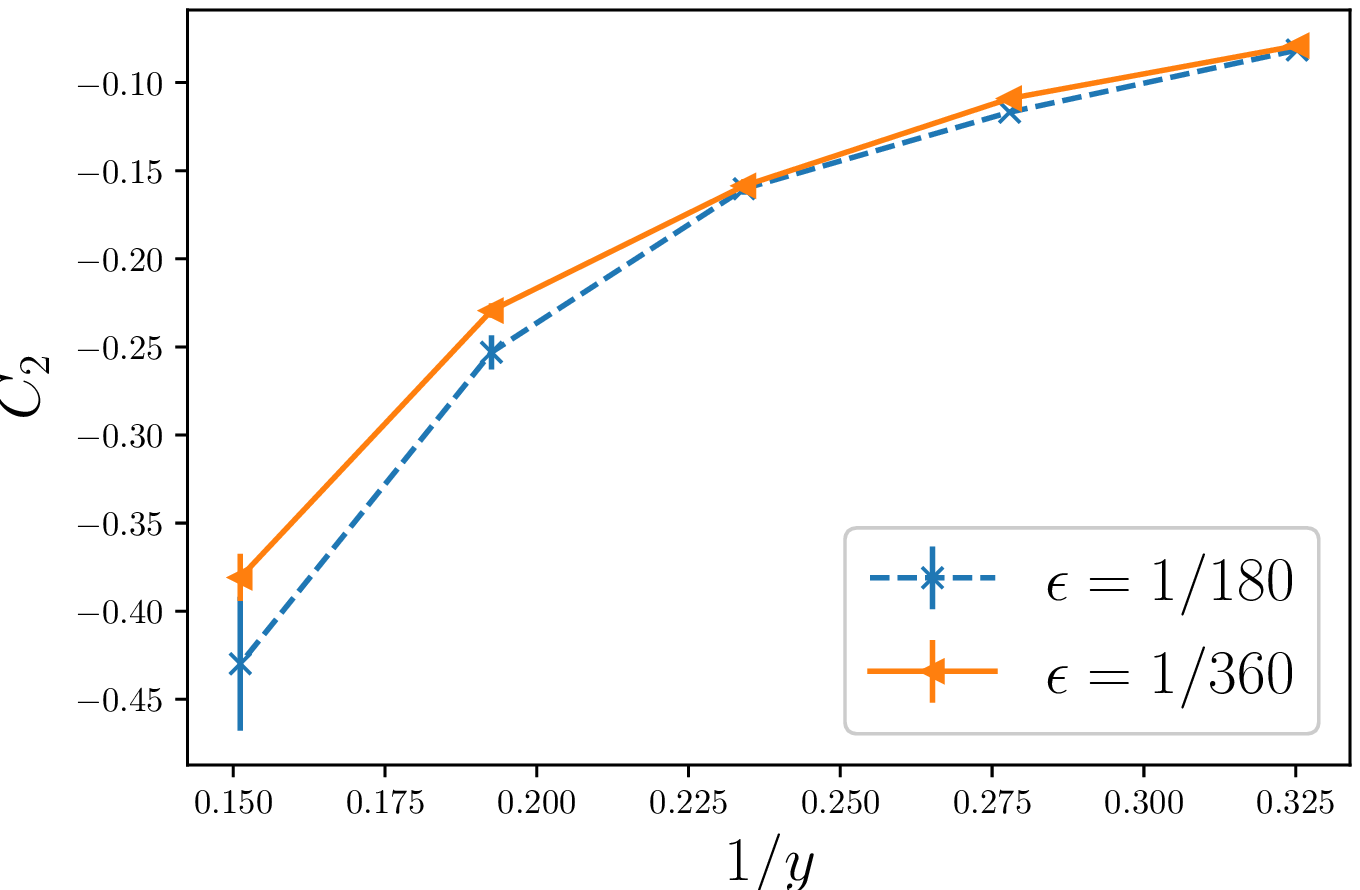}
	\includegraphics[scale=0.4]{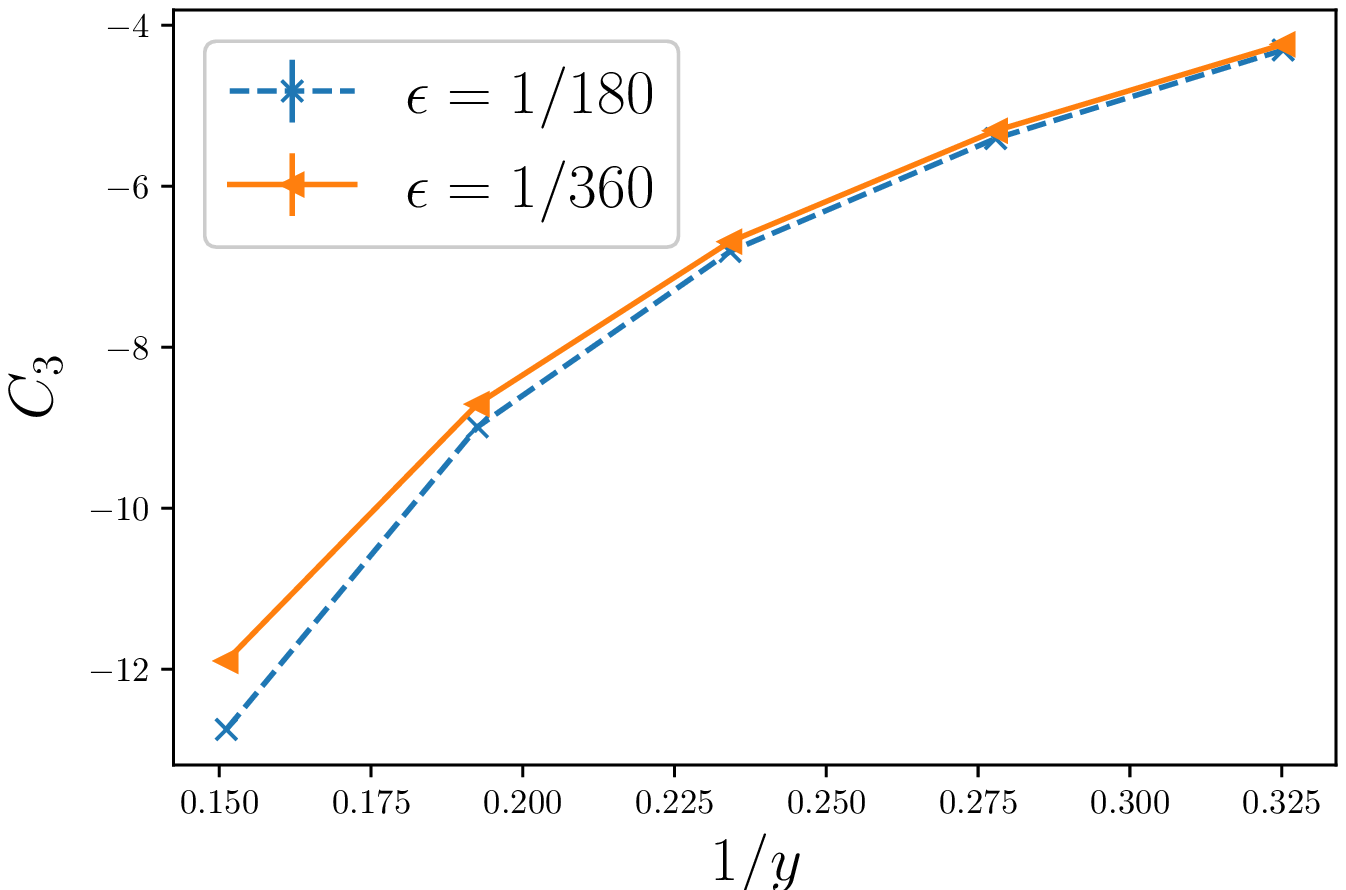}
	\includegraphics[scale=0.4]{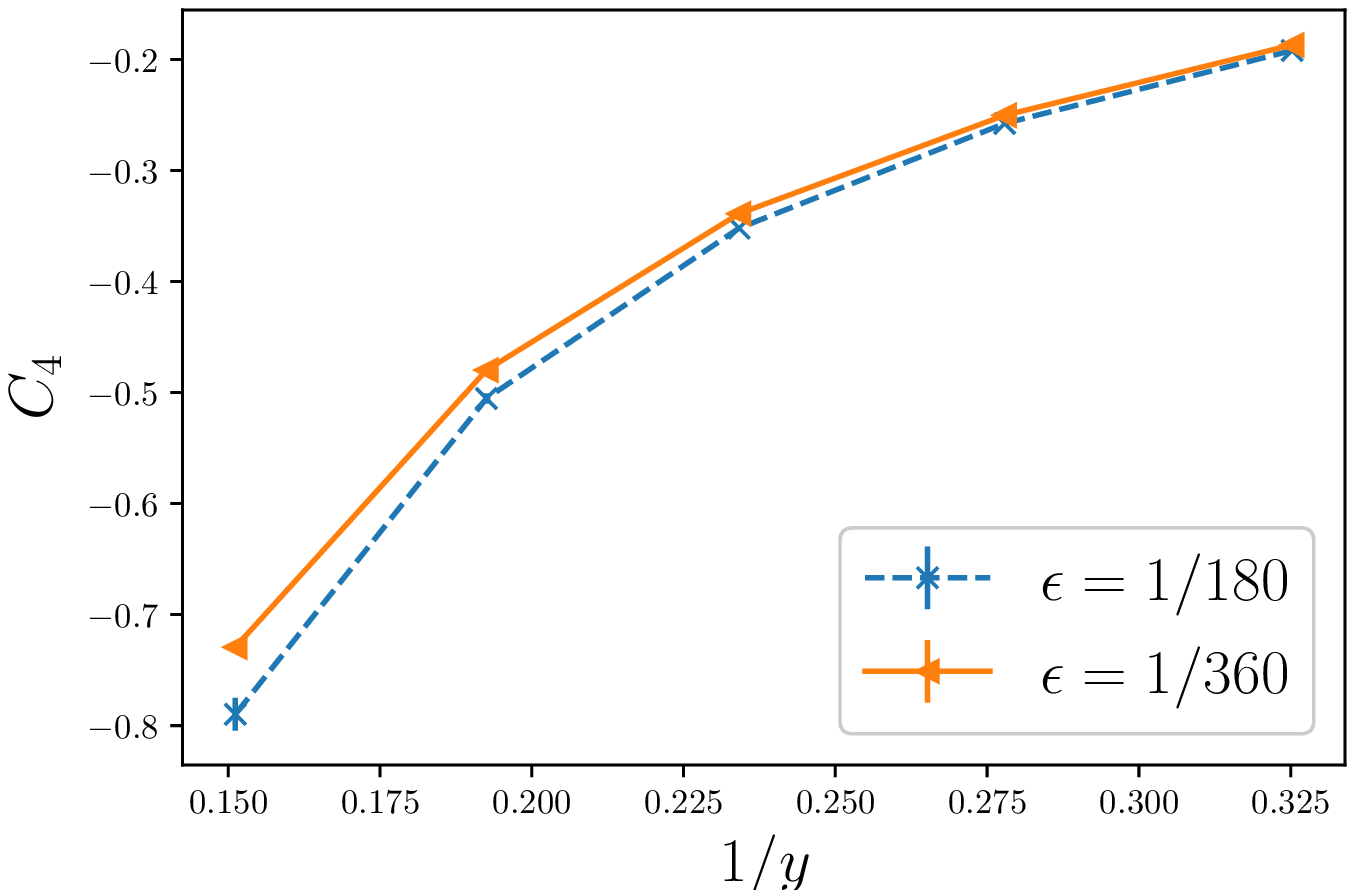}
	\includegraphics[scale=0.4]{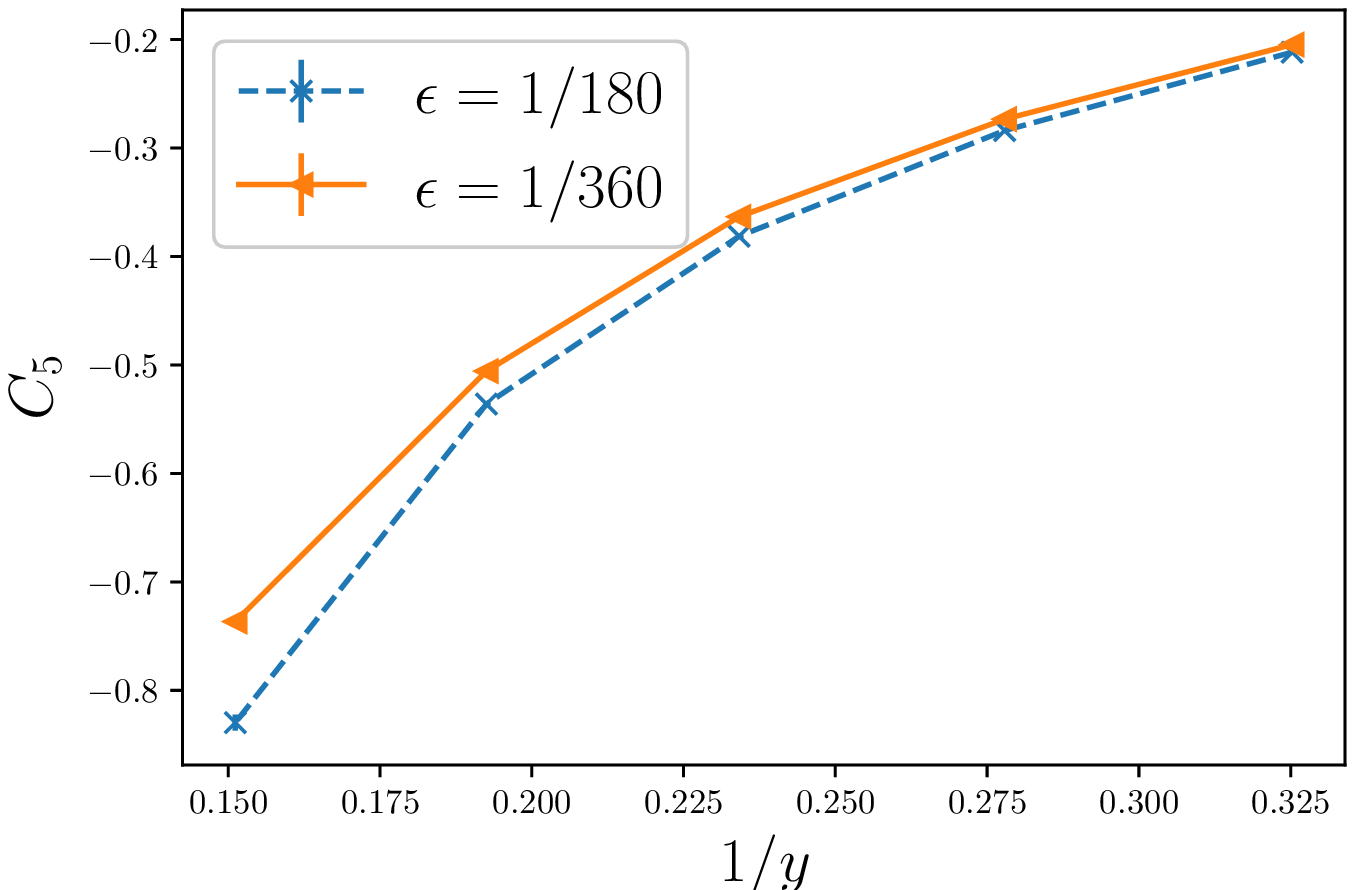}
	\includegraphics[scale=0.4]{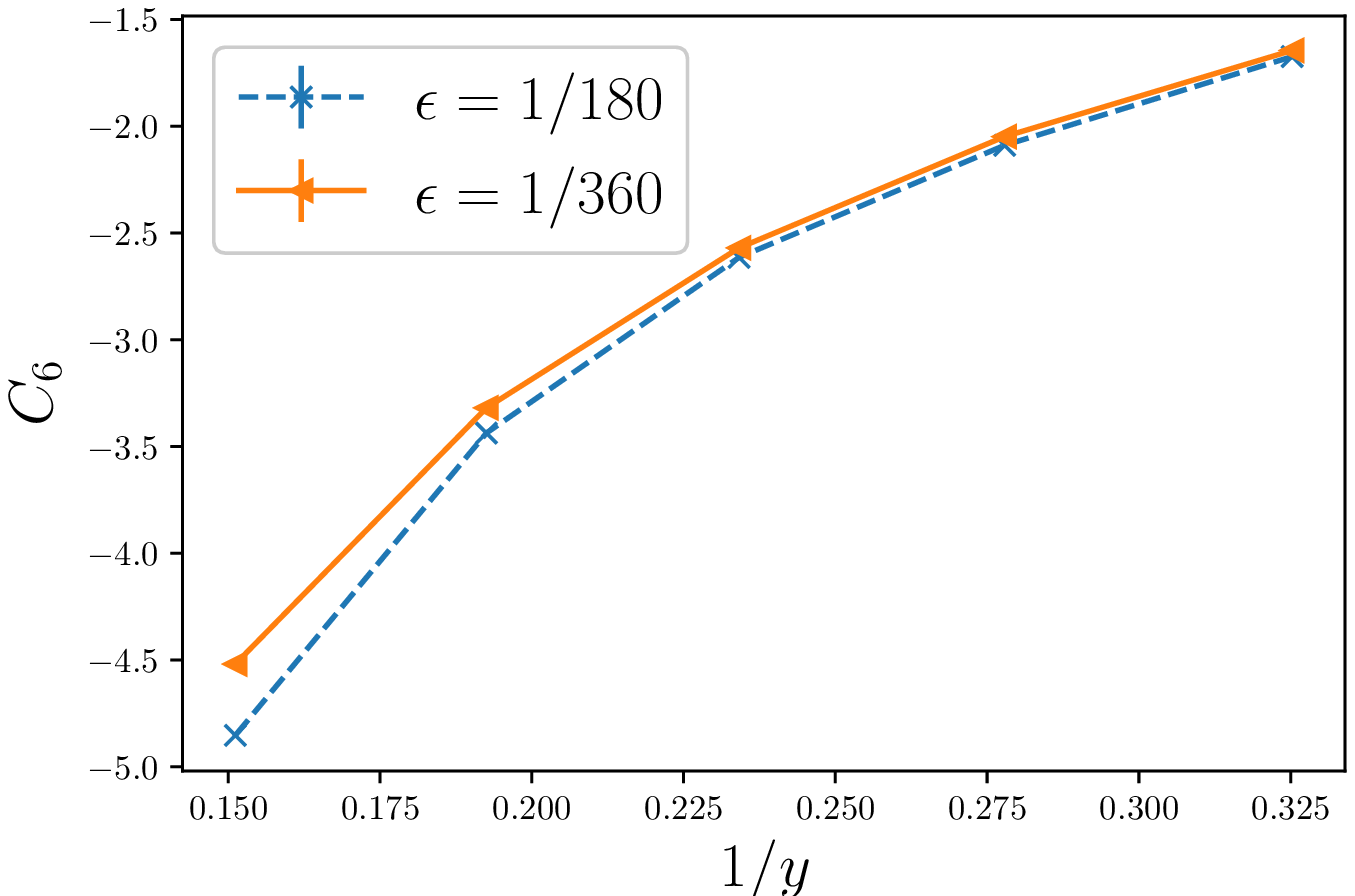}
	\caption{The figure shows the numerical solution of the integrals in Eq. \ref{20} in function of $1/y$ for different values of the discretization interval $\epsilon$.}
	\label{figure_C}
\end{figure}
In the generation of the trajectories the values of $y(q_i)$ and $m(q_i,z_i)$
are obtained through interpolation from a table of results for
 $q(x)$, $y(q)$ and $m(q,z)$ at $T=0$ calculated with the same method of \cite{charbonneau2014exact},  with 40 RSB. In this way we are able to calculate the integrals in Eq. \ref{20} at $T=0$ (see \ref{app:num_ev}).\\

\begin{figure}
	\centering
	\includegraphics[scale=0.5]{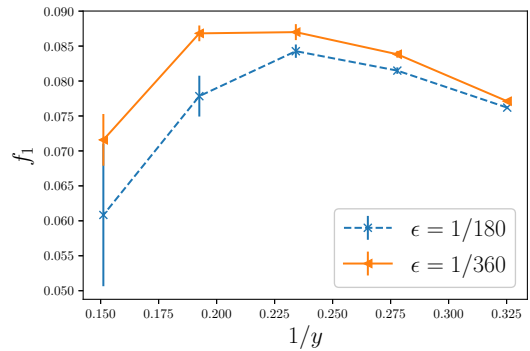}
	\caption{Numerical solution of $f_1$ in function of $1/y$ for different values of the discretization interval $\epsilon$.}
	\label{figure_f1}
\end{figure}

We will now focus on the behaviour of the divergent terms in function of $y$.
In Fig. \ref{figure_C} we plot the solution of the different integrals of Eq. \ref{20} in function of  $1/y \sim T$ for $\epsilon=1/180$ and $\epsilon=1/360$.  The figure shows that all the terms diverge for $T \to 0$. However, when we plot $f_1$ (Fig. \ref{figure_f1}) we can notice that there are great cancellations between the different terms. Looking at Fig. \ref{figure_f1} we can see that $f_1$ appears to have a "residual" divergence for $T\to 0$. This divergence can be addressed to the approximations done to compute numerically the integrals, which are the discretization of the differential equation in Eq. \ref{15} and the finite, even if large, number of replica symmetry breakings used to estimate $q(x)$. The effect of the finite size of $\epsilon$ is evident looking at Fig. \ref{figure_f1}, where we compare the values of $f_1$ for $\epsilon=1/180$ and $\epsilon=1/360$. For $\epsilon=1/360$ we can see that the divergence of $f_1$ tends to decrease as a result of a better cancellation between the different integrals. This improvement with the decreasing of $\epsilon$ suggests that for  $\epsilon \to 0$ the divergence disappears leaving a finite value for $f_1$.
Given the divergences due to these errors we couldn't calculate the values of $f_1$ at small T, we used a linear extrapolation to find its value at $T=0$. Using the last two estimated points $(1/y)_{1}=0.33$ e $(1/y)_2=0.28$ and their corresponding ordinates, we obtained:
\begin{equation}
f_1(T=0)= 0.13 \pm 0.01 \ .
\end{equation}
The error being purely statistical. Looking at the curve obtained by G.Parisi e F. Tria in \cite{parisi2002spin}, we can notice that in that case an extrapolation at $T=0$ would lead to a greater value respect to the one evaluated with our data. The reasons for this discrepancy can be attributed to the underestimation due to the finite step $\epsilon$ used to evaluate the trajectories in Eq. \ref{15}, that could bring to a systematic error of the same order of the statistical one.
The value we get for $f_1(T=0)$, i.e. $0.13\pm 0.01$ is not far from the numerical value of  $\approx 0.17$ found in \cite{boettcher2003numerical}.

We can conclude that there are analytical and numerical evidence that the expansion of the free energy around the point of infinite connectivity can be successfully computed at low temperature. Similar, but albeit more difficult, computations can be done in mean field model of structural glasses in high, but finite dimensions. \\

\appendix 

\section{ The definition of the $q$'s functions}
\label{appendix_qi}
The functions that appears in \ref{17}, corresponding to the trees in \ref{3}, are defined as
\begin{eqnarray}
	q_1(x_1,x_2,x_3)&=&q_{abcd}|_{a\wedge b=x_1, c\wedge d = x_2, a \wedge c=b \wedge c = a \wedge d = b\wedge d = x_3}\\
	q_2(x_1,x_2,x_3)&=&q_{abcd}|_{a\wedge b=x_1, a\wedge c = b \wedge c=x_2, a \wedge d = b \wedge d = c\wedge d = x_3}\\
    q_3(x_1)&=&q_{abcd}|_{a\wedge b= a\wedge c = a \wedge d = b\wedge c = b \wedge d = c \wedge d = x_1}\\    
    q_4(x_1,x_2)&=&q_{abcd}|_{a\wedge b= a\wedge c = x_1, a \wedge d = b\wedge d= c\wedge d=x_2}\\ 
    q_5(x_1,x_2)&=&q_{abcd}|_{a\wedge b= = x_1, a \wedge c = a\wedge d= b\wedge c = b\wedge d= c \wedge d =x_2}
\end{eqnarray}

\section{Details of the numerical evaluation of $f_1$}
\label{app:num_ev}

We want to evaluate integrals of the form of Eq. \ref{17}. Let us take the following one (corresponding to the second tree in Fig. \ref{3}) 
\begin{equation}
\int_{0}^{\beta}dy_3\int_{y_3}^{\beta}dy_2\int_{y_2}^{\beta}dy_1q_2(y_1,y_2,y_3)^2\ .
\end{equation}
We can rewrite it in the following way
\begin{equation}\label{integral_appendix}
\int_{0}^{\lambda}dq_3 \frac{dy_3}{dq_3}\int_{q_3}^{\lambda}dq_2\frac{dy_2}{dq_2}\int_{q_2}^{\lambda}dq_1\frac{dy_1}{dq_1}q_2(y_1,y_2,y_3)^2\ ,
\end{equation}
imposing a cut off $\lambda$ to control the divergence of the different terms for $\lambda \to 1$. The argument of the integral is described in Eq. \ref{q2}. In order to evaluate it we generate randomly $q_1,q_2\ \mbox{and} \ q_3$ in the interval $[0,1]$ a large number $M$ of times  such that $q_1>q_2$ e $q_2>q_3$. For each random generation of the three variables we calculate the argument of the integral in these points and than take the average over all the $M$ generations. For each generation $j$ we will have
\begin{equation}\label{21}
q_2^2(y_1,y_2,y_3)_j=m_{\eta}(q_3)_jm_{\eta}(q_2)_jm^2_{\eta}(q_1)_jm_{\omega}(q_3)_jm_{\omega}(q_2)_jm^2_{\omega}(q_1)_j\ ,
\end{equation}
such that
\begin{eqnarray}
&\int_{0}^{\beta}dy_3\int_{y_3}^{\beta}dy_2\int_{y_2}^{\beta}dy_1 q_2(y_1,y_2,y_3)^2 =\\ \nonumber &=\frac{1}{M}\sum_j^{M}m_{\eta}(q_3)_jm_{\eta}(q_2)_jm^2_{\eta}(q_1)_jm_{\omega}(q_3)_jm_{\omega}(q_2)_jm^2_{\omega}(q_1)_j\ .
\end{eqnarray}
 In order to evaluate $m_{\eta}(q)$ and $m_{\omega}(q)$ we need to take two random walks (with noises $\eta$ and $\omega$) using Eq. (\ref{15}) which starts from $q=0$, pass trough $q_3$ and then branches in two leaves ending at $q_1$ and $q_2$, with $q_1<\lambda$ e $q_2<\lambda$ (see figure \ref{3}). We should notice that it is sufficient to average the generation of the random overlaps, used to perform the integration, in order to perform also the average over the different trajectories. The same procedure has been applied to calculate the other integrals appearing in Eq. \ref{20}.


\section*{Aknowledgments}
This project has received funding from the European Research Council
(ERC) under the European Union's Horizon 2020 research and innovation
programme (grant agreement No [694925]).

\section*{References}

\bibliographystyle{ieeetr}
\bibliography{biblio}{}
\end{document}